\documentclass[twocolumn,showkeys,preprintnumbers,amsmath,amssymb]{revtex4}

\usepackage{graphicx}
\usepackage{dcolumn}
\usepackage{bm}

\begin{document}

\newcommand{\omegar}{\omega_{\rm rf}}
\newcommand{\omegas}{\omega_{\rm s}}
\newcommand{\omegag}{\omega_{\rm g}}

\preprint{APS/123-QED}

\title{Audio mixing in a tri-port nano-electro-mechanical device}

\author{M. Defoort$^1$}
\author{K. Lulla$^1$}
\author{J-S. Heron$^1$}
\author{O. Bourgeois$^1$}
\author{E. Collin$^1$} 
\email{eddy.collin@grenoble.cnrs.fr}
\author{F. Pistolesi$^2$}

\affiliation{%
$^1$Institut N\'eel, 
CNRS et Universit\'e Joseph Fourier, \\
BP 166, 38042 Grenoble Cedex 9, France \\
}%
\affiliation{%
$^2$Laboratoire d'Ondes et Mati\`ere d'Aquitaine, Universit\'e Bordeaux I \& CNRS, \\
351 cours de la Lib\'eration, 33405 Talence cedex, France \\
}%

\date{\today}

\begin{abstract}
We report on experiments performed on a cantilever-based tri-port nano-electro-mechanical (NEMS) device. Two ports are used for actuation and detection through the magnetomotive scheme, while the third port is a capacitively coupled gate electrode. 
By applying a low frequency voltage signal on the gate, we demonstrate mixing in the mechanical response of the device, even for {\it low magnetomotive drives, without resorting to conduction measurements through the NEMS}. The technique can thus be used in particular in the linear regime, as an alternative to nonlinear mixing, for normal conducting devices. An analytic theory is presented reproducing the data {\it without free parameters}.
\end{abstract}

\maketitle

Signal mixing is a very important tool of signal processing in conventional electronics.
The basic application consists in translating an audio signal (typ. 0.01 - 15 kHz) into the radio-frequency bandwidth (10 - 100 MHz), enabling thus a straightforward broadcasting and multiplexing.
Nowadays, nano-electro-mechanical (NEMS) devices based on silicon technologies routinely reach frequencies ranging from 1 up to 100 MHz \cite{someone1,someone2}, and can thus be thought of as potential components for electronic circuits.

The first works on nanomechanical mixing used an intrinsic nonlinearity of the device \cite{mix}. The NEMS under study had a normal metal conducting layer, and needed to be used in its nonlinear dynamic range, under large deflections.
Another scheme combines a gate capacitance driven at high frequencies close to resonance, with a semiconducting property of the device itself driven at low frequency \cite{McEuen}. This method proves to be extremely useful for detecting the motion itself of very high frequency nanotube-based devices, through the imprint of the mixing effect in the conductance \cite{vanderzant}. It has been developed since, with the adjunction of the field effect \cite{purcell} and the impressive achievement of the ``nano-radio'' \cite{radio}.

\begin{figure}[t!]
\includegraphics[height=5.5 cm]{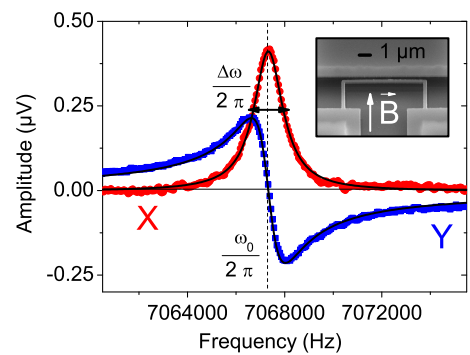}
\caption{\label{ref} 
(Color online) 
The $X$ and $Y$ components (dots) for the bare 
resonance line measured in vacuum at 4.2 K with a magnetic field $B$ of 840 mT
and a drive current amplitude $I_0=0.47$ $\mu$A (grounded gate voltage). 
The resulting displacement amplitude is $x_0= 5.6$ nm. 
Lines are Lorentzian fits, leading to $\omega_0/2\pi=7.067$ MHz (position) and $\Delta \omega /2\pi =1.45$ kHz (linewidth). 
Inset: SEM image of the sample with indicated the direction of the 
static magnetic field.}
\end{figure}

In this paper we present a capacitive mixing scheme which applies to normal conducting NEMS kept in their linear regime.
The device we use is the goal-post shaped silicon device presented in Ref.~\cite{qfsus}. 
It consists in two ``feet'' linked by a ``paddle'' of length $l \approx 7$ $\mu$m 
and thicknesses 150 nm, coupled to a large gate electrode. 
It has been fabricated from a thick SOI substrate, with the adjunction of a 30 nm aluminum layer on top. 
The drive and detection schemes use the magnetomotive technique: the paddle lies in a 
perpendicular, in plane,
static field $B$ while an oscillating current $I_0 \cos(\omegar t)$ is fed through it. 
An alternating force of amplitude $F_0=I_0 l B$, that we keep small, excites the first (out-of-plane) mode of the structure at an amplitude $x_0$,
when $\omegar \approx \omega_0$ ($\omega_0$ being the mode resonance frequency). 
The motion induces in turn a voltage $l B \omegar x_0 \cos(\omegar t + \phi)$ that we detect with a lock-in amplifier referenced at $\omegar$ ($X=l B \, \omegar x_0 \cos(\phi)$ and $Y=l B \, \omegar x_0  \sin(\phi)$ components). 
Experiments are performed at 4.2 K in cryogenic vacuum ($P < 10^{-6} $mbar).
In Fig. \ref{ref} we present a measured reference resonance line together with an SEM picture of the device. Note the quality of the Lorentzian fits proving excellent linearity.

On the gate electrode we apply a time-dependent voltage $V_g (t)=V_{g0} \cos(\omegas t)$, with $\omegas \ll \omegar$. 
This drive generates an additional out-of-plane force 
$
F_g (t) = (1/2) V_g(t)^2 \partial C(x)/\partial x  
$ 
on the moving structure, $C$ being the total capacitance between the NEMS and the fixed electrode. This force is highly nonlinear, and at first order $\partial C(x)/\partial x \approx \partial C(x=0)/\partial x + [ \partial^2 C(x=0)/\partial x^2 ] \,x$. We have extremely carefully calibrated our device, demonstrating in particular that the first $\partial C(x=0)/\partial x$ term, which only creates a static shift in the structure's rest position, can be safely neglected \cite{JAP}. 
The equation of motion thus takes the form:
\begin{equation}
  m \, \ddot{x}+ 2 \Lambda \, \dot{x} +
	\left[ k + \delta k \left(1+ \cos[2 \omegas t]\right) \right] \, x 
 =  F_0 \cos(\omegar t)  
, \label{dyn}
\end{equation}
where $ \delta k  = - ({1}/{4}) V_{g0}^2 \,  {\partial^2 C (x=0)}/{\partial x^2}$,
$m$, $k$, and $\Lambda$ are the normal mass, the spring constant, and the damping constant
of the mode under study, respectively.
By definition $\omega_0=\sqrt{k/m}$ is the ``bare'' resonance frequency and $\Delta \omega = 2 \Lambda / m$ the full linewidth at half height on the lock-in $X$ component (both in Rad/s).

The problem at hand is a forced Mathieu equation. 
The homogeneous case ($F_0=0$) has been studied in details in the literature \cite{NafiafBook}. 
To find the stationary solution of the inhomogeneous equation (\ref{dyn}) 
we introduce an {\em ansatz} for $x(t)$:
\begin{equation}
	x(t)  
	=  
	\sum_{n=-\infty}^{+\infty} 
	|A_n| \cos(\omegar t + 2n \omegas t + {\rm arg}(A_n))
	\,,
 	\label{xmix}
\end{equation} 
with $A_n$ complex coefficients. 
Substitution of Eq.~(\ref{xmix}) into Eq.~(\ref{dyn}) gives the following 
recurrence equation:
\begin{equation}
a_n(\omegar) A_n+ A_{n+1}+A_{n-1}
=  \delta_{n,0} \, 2 F_0/\delta k
\,,
\end{equation}
where we introduced:
$a_n(\omegar)={\cal L}(\omegar + 2n \omegas) 2 (k+ \delta k)/\delta k$,
with
${\cal L}(\omega)=(\omegag^2-\omega^2- i  \omega \, \Delta \omega)/\omegag^2$,
and
$\omegag^2=(k+\delta k)/m$ the renormalized resonating frequency squared.
Here $\delta_{i,j}$ is the Kronecker's symbol.
Our measurement scheme allows to measure the response of the device at 
the $\omegar$ tone, but as it is clear from the form of $x(t)$, many other 
harmonics are present.
What is obtained from the measurement of $X$ and $Y$ 
is the real and imaginary parts of $A_0$.
An explicit expression of $A_0$ can be given in terms of continued
fractions \footnote{We use the standard notation for continued fractions
$\frac{1}{a_0-}\frac{1}{a_1-}\cdots = \frac{1}{a0-\frac{1}{a_1-\cdots}}$.}:
\begin{equation}
	A_0 = 
	\frac{2 F_0/ \delta k}
	{a_0
	-\frac{1}{a_1 -}\frac{1}{a_2 -}\frac{1}{a_3 -}\cdots
	-\frac{1}{a_{-1} -}\frac{1}{a_{-2} -}\frac{1}{a_{-3} -}\cdots
	}
	\,.
\label{soluce}
\end{equation}
With no mixing ($\delta k \rightarrow 0$), Eq.~(\ref{soluce}) reduces to the standard 
harmonic Lorentzian solution $A_0 =2 F_0/a_0 \delta k=
F_0/[ m \, \omegag^2 {\cal L}(\omegar) ]$.
For non-vanishing $\delta k$ the effect of the mixing is to induce sidebands in the 
response function at intervals of frequency $2\omegas/2 \pi$.
\begin{figure}[t!]
\includegraphics[height=5.5 cm]{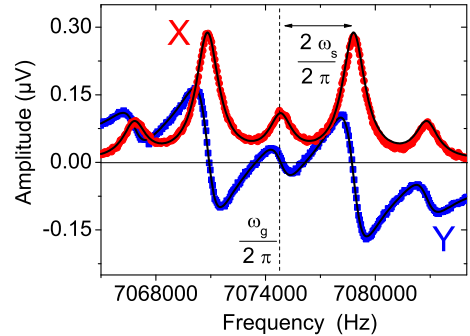}
\caption{\label{mixHF} (Color online) Measurement of the $X$ an $Y$ 
components (red and blue dots) with a magnetomotive drive of 
$I_0=0.93$ $\mu$A and $B=840$ mT in presence of 
a gate modulation of frequency $\omega_s=2$ kHz and 
voltage $V_{g0}=$ 1.4 V (implying $\delta k/k = 2. \times 10^{-3}$). 
The lines are our theoretical predictions.
}
\end{figure}
\begin{figure}[t!]
\includegraphics[height=5.5 cm]{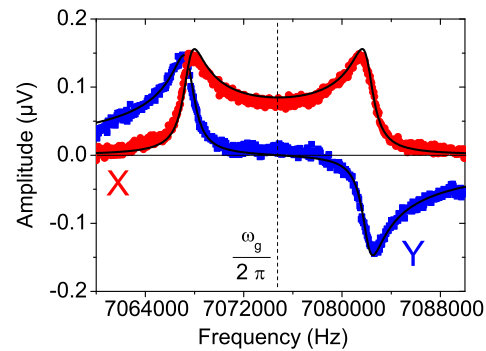}
\caption{\label{mixLF} (Color online) Same as Fig.~\ref{mixHF} with 
$\omega_s=5$ Hz.
}
\end{figure}

In Fig.~\ref{mixHF} we present the measured data for a gate modulation signal $V_{g0}$ of 1.4 V and a frequency of 2 kHz. The magnetomotive drive is kept in the linear regime. The lines are calculated numerically from Eq. (\ref{soluce}) taking into account 50 nested fractions 
in the recursion (each fraction adds one pole at positive frequency to the response function). 
We found however that for these values of the parameters adding more than 5 terms in the recursive expression 
does not change visually the result.
All numerical parameters ($m$, $\partial^2 C (x=0)/\partial x^2$, etc...) are obtained 
from the careful calibration of the setup \cite{JAP}: there are thus no free parameters 
in the theoretical prediction. 
The agreement is excellent, and one can clearly distinguish in the $X$ signal 
the peaks separated by $2 \omegas/2 \pi$. 

The behavior presented in Fig. \ref{mixHF} is representative for values of $\omegas \gg \Delta \omega$.
In the opposite limit we found that the peaks in the $X$ signal merge together to 
form a characteristic structure of hat shape.
The experimental results are shown in Fig.~\ref{mixLF}, where 
the gate modulation frequency $\omegas/2\pi=5$ Hz.
Also in this case the theory works perfectly well, but one needs to include 50 nested 
fractions to reach convergence. 
A simple solution can be actually found in this limit.
The gate modulation is so slow that the oscillator reaches the stationary 
oscillation state before the gate changes of an appreciable quantity.
In practice focusing on the $X$ signal one would see a 
lorentzian peak that bounces back and forth in the frequency 
plot around $\omegar$ with an amplitude of $\delta k/2\pi m$.
Since the measurement averages over a time longer than
$1/\omegas$ (typ. 1 s) the result is this hat shaped plot shown 
in Fig.~\ref{mixLF}.
An analytical expression for this can be readily found by averaging
the quasi-stationary value of $A_0$ over a period $2\pi/\omegas$. 
This gives: 
\begin{equation}
	A_0 = {F_0\over m\omegag^2} 
 	\int_0^{2\pi/2\omegas}\!\!\!\!\!\!\!\!\!\!\!\!\!\!\!\!\!
	\frac{(2 \omegas/ 2 \pi) \, dt }{{\cal L}(\omegar) +
\frac{\delta k}{k+\delta k} \cos(2\omegas t)}
=
\frac{F_0}{m \omegag^2} f \left(\frac{a_0}{2} \right)
\,,
\end{equation}
with 
$f(z)={\rm sign}({\rm Re}[z])/\sqrt{z^2-1}$.
We found that this expression coincides with the exact expression,
Eq. (\ref{soluce}), with 50 nested fractions for the case of Fig. \ref{mixLF}.

\begin{figure}[t!]
\includegraphics[height=5.5 cm]{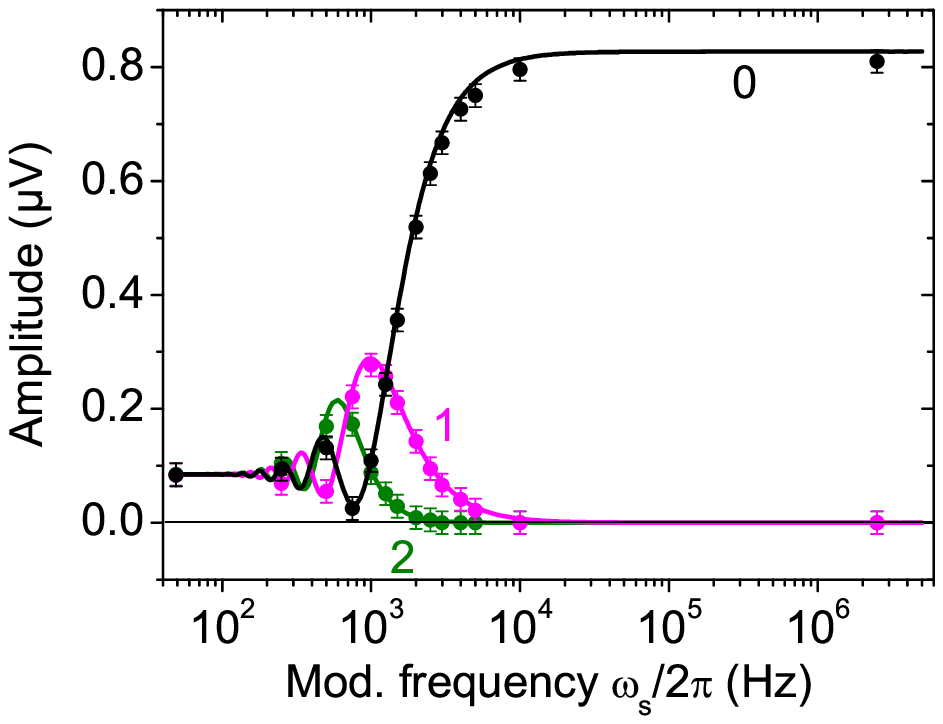}
\caption{\label{freqdep} (Color online) 
Measured (dots) and predicted (lines) heights of the first poles as a function of the modulation frequency $\omegas/2\pi$ (same parameters as Fig.~\ref{mixHF}).
}
\end{figure}

\begin{figure}[t!]
\includegraphics[height=5.5 cm]{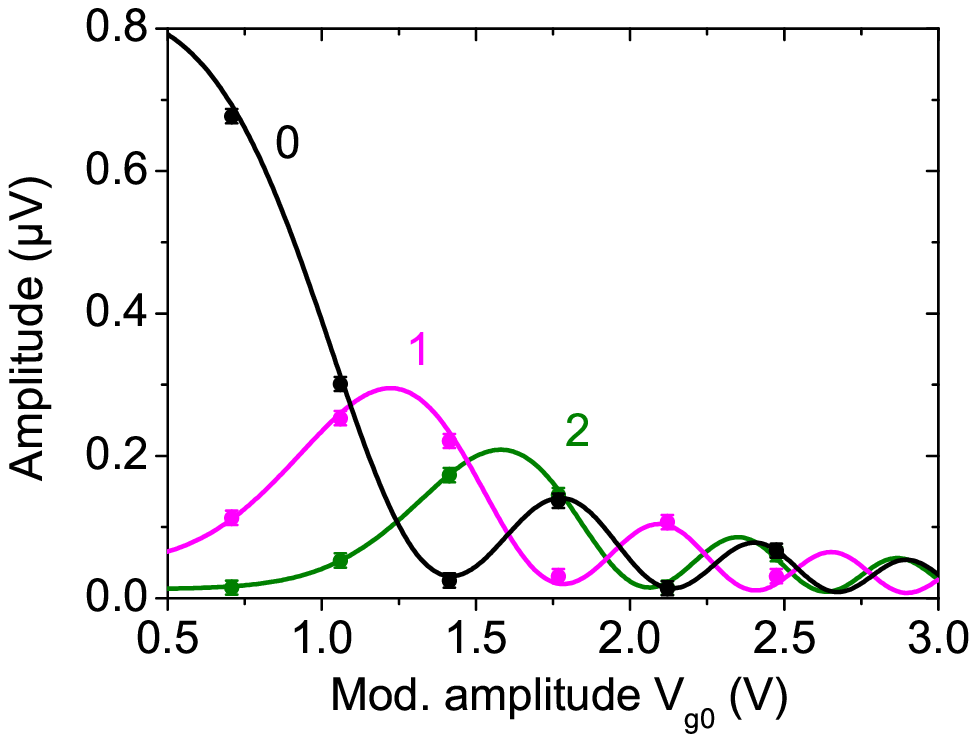}
\caption{\label{drivedep} (Color online)
Measured (dots) and predicted (lines) heights of the first poles as a function of the modulation amplitude $V_{g0}$ (same parameters as Fig.~ \ref{mixHF}).
}
\end{figure}

We finally illustrate the overall behavior on Figs. \ref{freqdep} and \ref{drivedep}. In both graphs we plot the height of the first poles, defined as the amplitude of the signal measured at $\omegag \pm 2 n \omegas$ ($n=0,1,2$). In Fig. \ref{freqdep} we present the pole height dependence as a function of the modulation frequency (fixed $V_{g0}=1.4$ V), while in Fig. \ref{drivedep} we present the modulation amplitude dependence (fixed $\omegas/2\pi$=1.5 kHz). The lines are our theoretical predictions, with no free parameters, proving excellent agreement.


In conclusion, we presented a mixing technique that {\it does not} require a nonlinear drive of the nanomechanical device. The measure is performed with the standard magnetomotive technique, and {\it does not} require transport measurements to be performed. We present the analytic theory describing the results. Data are reproduced {\it without free parameters}, using the NEMS carefully measured characteristics.
We demonstrate excellent understanding of the NEMS dynamics, and believe that our results will lead to future works in nanomechanical mixing.

\begin{acknowledgments}
We wish to thank T. Fournier, C. Lemonias, and B. Fernandez for their help in the fabrication of samples, and  
V. Bouchiat for valuable discussions.
We acknowledge the support from MICROKELVIN, the EU FRP7 low temperature infrastructure grant 228464, and of the 2010 ANR French grant QNM n$^\circ$ 0404 01.
\end{acknowledgments}



\begin{thebibliography}{0}
\expandafter\ifx\csname natexlab\endcsname\relax\def\natexlab#1{#1}\fi
\expandafter\ifx\csname bibnamefont\endcsname\relax
  \def\bibnamefont#1{#1}\fi
\expandafter\ifx\csname bibfnamefont\endcsname\relax
  \def\bibfnamefont#1{#1}\fi
\expandafter\ifx\csname citenamefont\endcsname\relax
  \def\citenamefont#1{#1}\fi
\expandafter\ifx\csname url\endcsname\relax
  \def\url#1{\texttt{#1}}\fi
\expandafter\ifx\csname urlprefix\endcsname\relax\def\urlprefix{URL }\fi
\providecommand{\bibinfo}[2]{#2}
\providecommand{\eprint}[2][]{\url{#2}}

\end{thebibliography}


\begin{thebibliography}{blaaaaaaaaaaaaaaaaaaaaaaaaaaaaaaaaaaaaaaaaaaaa}

\bibitem{someone1} J. S. Aldridge and A. N. Cleland, {\it Phys. Rev. Lett.} {\bf 94}, 156403 (2005). 
\bibitem{someone2} R. B. Karabalin, X. L. Feng, and M. L. Roukes, {\it Nano Letters} {\bf 9}, 3116 (2009).
\bibitem{mix} A. Erbe, H. Kr\"ommer, A. Kraus, R. H. Blick, G. Corso and K. Richter {\it Appl. Phys. Lett.} {\bf 77}, 3102 (2000); A. Erbe and R. Blick, {\it IEEE Transactions on Ultrasonics, Ferroelectrics, and Frequency Control} {\bf 49} 1114 (2002).
\bibitem{McEuen} Vera Sazonova, Yuval Yaish, Hande Ust\"unel, David Roundy,
Tom\'as A. Arias \& Paul L. McEuen, {\it Nature} {\bf 431}, 284 (2004).
\bibitem{vanderzant} Benoit Witkamp, Menno Poot, and Herre S. J. van der Zant, {\it Nano Letters} {\bf 6}, 2904 (2006).
\bibitem{purcell} S. T. Purcell, P. Vincent, C. Journet et V. T. Binh, {\it Phys. Rev. Lett.} {\bf 89}, 276103 (2002).
\bibitem{radio} K. Jensen, J. Weldon, H. Garcia and A. Zettl, {\it  Nano Letters} {\bf 7}, 3508 (2007).
\bibitem{qfsus} E. Collin, T. Moutonet, J.-S. Heron, O. Bourgeois, Yu.M. Bunkov, H. Godfrin, {\it J. of Low Temp. Phys.} {\bf 162},  653 (2011).
\bibitem{JAP} E. Collin {\it et. al}, submitted to {\it J. of Applied Phys.} (2011).
%
\bibitem{NafiafBook} A. H. Nayfeh and D. T. Mook, {\em Nonlinear oscillations},  Wiley (1995).
\end{thebibliography}
\end{document}